\documentclass{article} % For LaTeX2e
\usepackage{nips13submit_e,times}
\usepackage{hyperref}
\usepackage{url}
\usepackage{graphicx}
\usepackage{amsmath}
\usepackage{amsfonts}
\usepackage[scriptsize]{subfigure}
\usepackage{float}
\usepackage{algpseudocode}

\title{Probabilistic Programming for Malware Analysis}

\author{
Brian Ruttenberg \\
Charles River Analytics\\
625 Mt. Auburn St\\
Cambridge, MA, 02138 \\
\texttt{bruttenberg@cra.com} \\
\And
Lee Kellogg \\
Charles River Analytics\\
625 Mt. Auburn St\\
Cambridge, MA, 02138 \\
\texttt{lkellogg@cra.com} \\
\And
Avi Pfeffer \\
Charles River Analytics\\
625 Mt. Auburn St\\
Cambridge, MA, 02138 \\
\texttt{apfeffer@cra.com}
}

\nipsfinalcopy % Uncomment for camera-ready version

\begin{document}

\maketitle

%\begin{abstract}
%Constucting lineages of malware is an important cyber--defense task. Performing this task is difficult, however, due to the amount of malware data and obfuscation techniques by the authors. In this work, we formulate the lineage task as a probabilistic model, and use a novel probabilistic programming solution to jointly infer the lineage and creation times of families of malware.
%\end{abstract}

\vspace{-5mm}
\section{Introduction}

Many malware authors borrow source code from other authors when creating new malware, or will take an existing piece of malware and modify it for their needs. As a result, malware within a family of malware (i.e., malware that is closely related in function and structure) often exhibit strong parent--child relationships. Determining the nature of these relationships within a family of malware can be a powerful tool for cyber--defense. 

Generating the \textit{lineage} of a family of malware is thus an important task but is difficult to perform manually due to the sheer volume of malware, intentional obfuscation by malware authors, and the many features and subtleties that must be examined to determine parent--child relationships. In this work, we describe a novel method for generating the lineage of a large family of obfuscated malware. We formulate the problem as a generative probabilistic model and develop a probabilistic programming (PP) algorithm to learn and infer the temporal and structural organization of a family's lineage. We demonstrated the accuracy and validity of our approach on synthetic data and real lab--generated malware. This work has significant implications in the cyber--defense community, and presents a problem that would be difficult to solve without the benefits of PP.

%intentionally obfuscate their code to deter analysis of provenance. Finally, there are many variables and subtleties in malware code that need to be examined to infer parent--child relationships. As such, generating the lineage of a family of malware is a task well--suited for automated artificial intelligence and machine learning methods.

\vspace{-2mm}
\section{Lineage as a Probabilistic Model}

The lineage of a set of malware binaries is a directed graph, where the nodes are the set of binaries in the family, and an edge from binary $A$ to $B$ implies that binary $B$ evolved partly from $A$ (and by implication, was created at a later time). The lineage can have multiple roots and binaries that contain multiple parents. By the definition of lineages, it is essential to know the order in which binaries were created. Without this information, it would be difficult to determine the inheritance direction of any parent--child relationships As such, the lineage of a set of binaries (i.e., the graph) is conditioned upon the the creation times of each of the binaries. We initially represent this simple relationship as high--level probabilistic model.

The \textit{Lineage} variable represents a distribution over the possible lineages that can be constructed from a set of binaries, conditioned upon the \textit{Creation Times} and the \textit{Malware Features}. The \textit{Creation Times} represents a distribution over the creation times of each binary and \textit{Malware Features} is a deterministic variable that constrains the lineage generation based on binary similarity. The more features that malware binaries share, the more likely they are connected in the lineage, but the actual parent--child assignment of the two nodes depends upon the given creation times. Lineage on a set of malware $\mathbb{M}$ is then defined as
\begin{equation}
\label{lineageMax}
Lineage_\mathbb{M} = \underset{Lineage_{\mathbb{M},i}}{argmax}\,\,P(Lineage_{\mathbb{M},i} | Features_\mathbb{M}, Times_\mathbb{M})
\end{equation}
Computing Eqn.~\ref{lineageMax} on a set of malware is often difficult because the compiler time stamp is often purposely obfuscated by the author, so the creation times must be inferred using any available information, either from within a binary or using external information. Fortunately, we can also use the date that malware was first encountered in the wild as additional evidence. 

One of the key insights is that the lineage and creation times are joint processes that can inform each other; knowing the lineage can improve inference of the creation times, and vice--versa. As such, performing joint inference of these models can potentially produce better results than inferring the creation times first and conditioning the lineage on the most likely creation times. 

\vspace{-1mm}
\subsection{Creation Time Model}

\begin{figure}[t]
\centering 
%\subfigure[Lineage as a random variable conditioned upon the creation times of the malware and their features.]{
%\includegraphics[width=.45\columnwidth]{Figures/LineageProbModel.pdf}
%\label{lineageprob}
%}
\subfigure[Plate for the model of each binary's creation time to infer a lineage--independent distribution of its creation time using time stamp and time seen as evidence.]{
\includegraphics[width=.45\columnwidth]{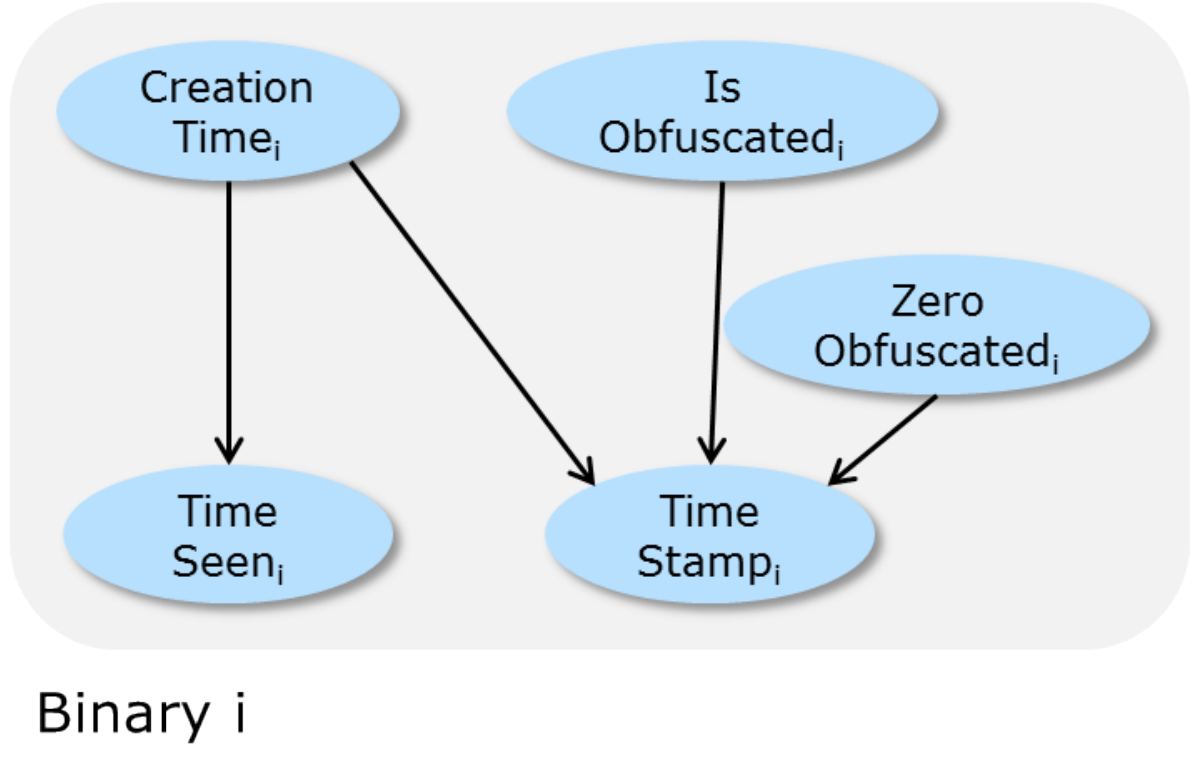}
\label{CreationTimeModel}
}
\subfigure[Lineage model given distributions of the creation times of each malware. The malware features are used as soft constraints on the inheritence relationships.]{
\includegraphics[width=.45\columnwidth]{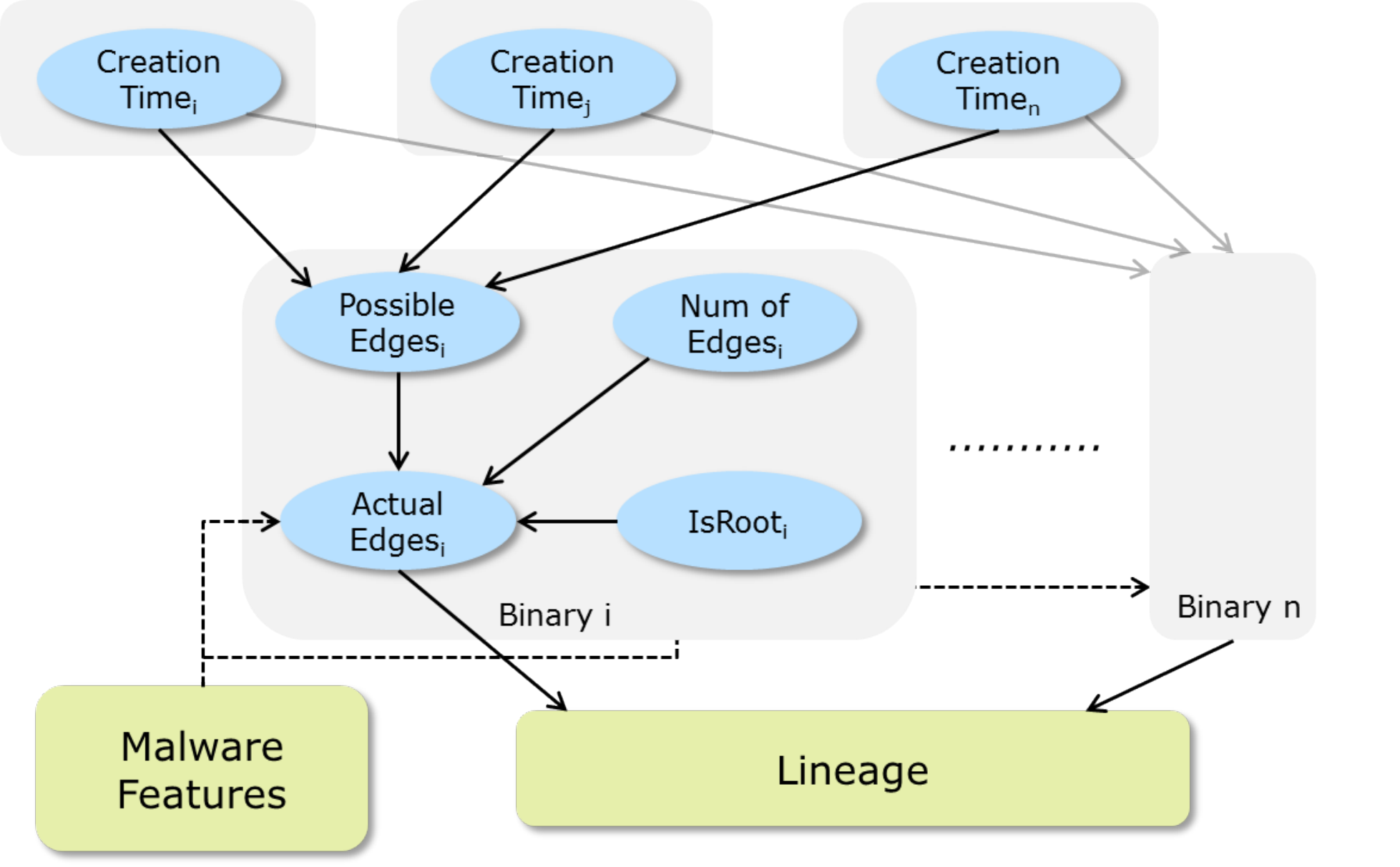}
\label{LineageModel}
}
\vspace{-2mm}
\caption[]{Probabilistic models of lineage}
\vspace{-5mm}
\end{figure}

The plate for the creation time model is shown in Fig.~\ref{CreationTimeModel}. For a set of $N$ binaries, we instantiate $N$ independent models. While the binary creation times are \textit{not} truly independent, the dependence between the creation times of different binaries is enforced through a joint inference algorithm, detailed in subsequent sections.

Each probabilistic model contains five variables. First, there is a variable to represent the actual binary creation time. There is also a variable to represent the time the binary was first seen in the wild. There is also a variable to represent the time stamp of the binary (from the actual binary header). This variable depends upon the creation time, as well as two additional variables that represent any obfuscation by the malware author to hide the actual creation time; one variable determines if the time stamp is obfuscated, and the other represents how the time stamp is obfuscated (either empty or some random value). Evidence is posted to the time seen and time stamp variables and a distribution of each malware's creation time can be inferred. Note that the priors for the obfuscation variables and parameters for the conditional distributions can be learned.

\vspace{-1mm}
\subsubsection{Lineage Model}

The model for lineage is shown in Fig.~\ref{LineageModel}. For each malware binary, we create a set of variables that represent how the binary can be used in the lineage. First, there is a variable which represents the possible existence of edges between a binary $i$ and all the other binaries. This variable is deterministic conditioned upon the creation times of all the binaries%; since a malware binary can only inherit from prior created binaries, the value of this variable is simply all of the binaries with earlier creation times.  

There are also two variables that control the number of edges (i.e., parents) for each binary, as well as a variable that specifies whether a particular binary is a root in the lineage. Finally, there is a variable that represents a set of \text{actual} lineage edges of a binary $i$ which depends upon the possible edges of the binary, the number of edges it has, and whether it is a root binary. By definition, the values of the actual edges variable for all binaries defines the lineage over the set of malware (i.e., it can be deterministically constructed from the edges and the creation times.)

The conditional probability distribution of the actual edges variable is constrained by the difference between the features of the binaries. That is, the higher similarity between two binaries, the more likely they are to have an edge between them. The similarity measure between binaries is based on binary similarity measures and we refer the reader to \cite{lakhotia_fast_2013} for more details.

\vspace{-2mm}
\section{Inference Algorithm}

As shown in Eqn.~\ref{lineageMax}, the lineage is the maximal probability lineage given the creation times and the malware features. Since the creation times are unknown, we must infer both the lineage and the creation times jointly. To accomplish this, we employed an iterative algorithm to jointly infer the most likely binary creation times and lineage, outlined as follows:

\begin{enumerate}
  \item \textbf{Infer a distribution of the binary creation times.} Using the observable time stamp and time seen information, we infer a distribution of the creation times of each binary. This distribution is still conditioned upon the lineage; this process marginalizes out some of the information not needed to compute a lineage. 
  \item \textbf{Sample the creation times.} We take a sample from the creation time distributions of the malware binaries. This creates a fixed order of the binaries.
  \item \textbf{Infer the most likely lineage.} We infer the most likely lineage of the malware binaries given the fixed creation times. That is, we compute the lineage described in Eqn.~\ref{lineageMax}.  
  \item \textbf{Infer the most likely creation times.} The most likely creation times is defined as the set of creation times that maximizes the probability of the creation times conditioned on the features and the current lineage.
%  \begin{equation}
%    Times_{\mathbb{M}} = \underset{Times_{\mathbb{M},i}}{argmax}\,\,P(Times_{\mathbb{M},i} | Features_\mathbb{M}, %Lineage_\mathbb{M})
%  \end{equation}
  Since we are conditioning on the previously computed lineage, we fix the inheritance between two binaries, but the direction of the edge can change depending on the inferred creation times.
  \item \textbf{Repeat steps 3--4 until convergence.} We repeat the process until convergence.
  \item \textbf{Repeat steps 2--5 until enough samples have been collected.} Since there is no guarantee that the maximization process converges on the global maximum, we restart the process to increase the likelihood that the global maximum is reached. %Let $Lineage^j_M$ and $Time^j_M$ be the lineage and creation times computed on the $j^{th}$ restart of the maximization process. Then the lineage and creation times returned to the user is
%  \begin{equation*}
%    (Lineage_M, Times_M) = \underset{Lineage^j_M, Times^j_M}{argmax}\,\,P(Times^j_M, Lineage^j_M | Features_M)
%  \end{equation*}
\end{enumerate}

The algorithm is very similar to the expectation--maximization algorithm~\cite{dempster1977maximum}, but we must re--sample several initial malware creation times to ensure that we are finding a satisfactory maximum value. At the end of the algorithm, we select the lineage with the highest probability (from the multiple restarts of the algorithm) as the lineage of the set of malware. %At each iteration, we find the most probable lineage and then subsequently the most probable creation times. So every iteration of the algorithm increases the joint probability of both lineage and creation times. 

\section{Implementation and Testing}

\begin{figure}
\centering 
\subfigure[Reduction in Error after Lineage.]{
\includegraphics[width=.45\columnwidth]{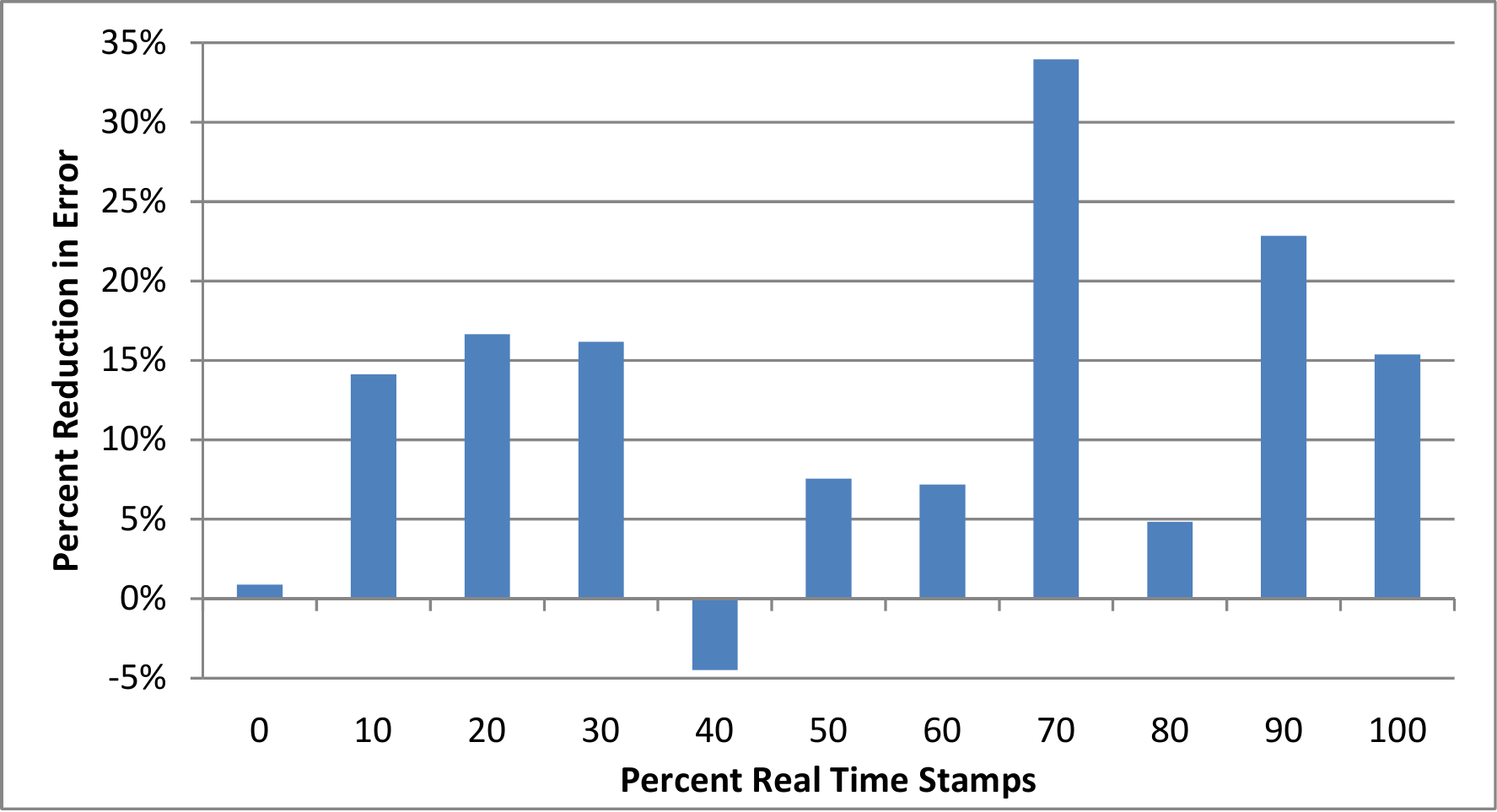}
\label{LineageTimeImprove}
}
\subfigure[Results of the lineage estimation on the MITLL data.]{
\includegraphics[width=.45\columnwidth]{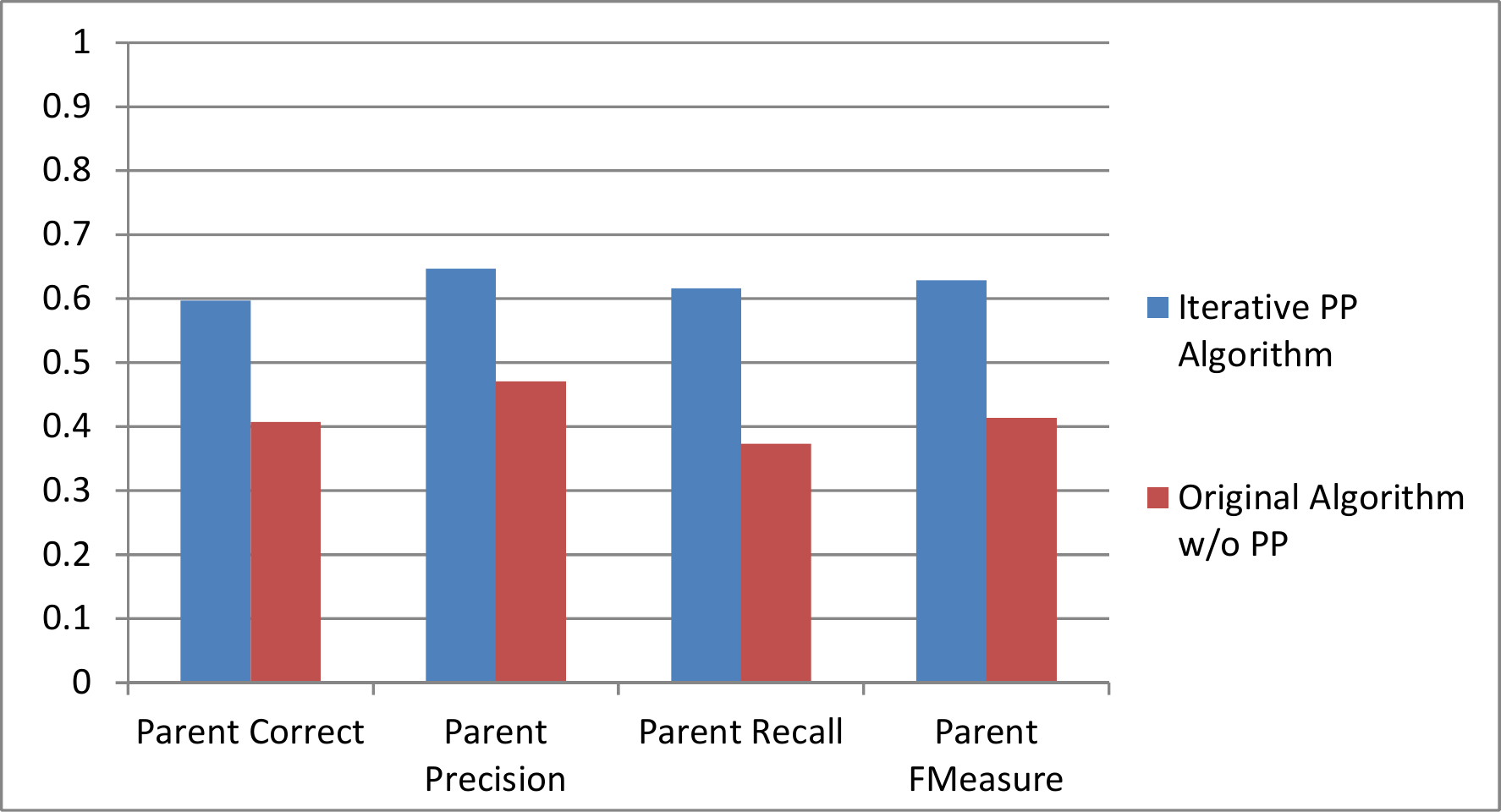}
\label{LineageResults}
}
\vspace{-3mm}
\caption[]{Testing results}
\vspace{-3mm}
\end{figure}

%\begin{figure}[t]
%\centering 
%\includegraphics[width=.65\columnwidth,height=50mm]{Figures/lineage-vis-example5.png}
%\vspace{-3mm}
%\caption{Interactive lineage algorithm implemented using probabilistic programming.\label{vis}}
%\end{figure}

We implemented the model in Figaro~\cite{figaro2012}, an open--source PP language, which provided several benefits on this problem. First, many of the model structures are repeated, so Figaro's object--oriented nature facilitated easy creation and reuse of these structures. Second, Figaro is a natural environment to create the iterative maximization algorithm; Figaro's inference engine can reason on any model encoded in the language, so it is easy apply successive maximization algorithms to the model. Finally, Figaro has several built in algorithms for learning and maximization, allowing us to focus on the model representation and not on the inference. %We also developed an interface to the lineage algorithm so that users can observe the iterative algorithm as it converges, and allow them to apply evidence of known inheritance relationships. An example of the interface is shown in Fig.~\ref{vis}.

Inferring the creation times depends upon variable parameters which are generally unknown. For instance, we need to know the prior probability that a binary's time stamp has been obfuscated. To determine these parameters, we learn them on a training set of data. Standard Bayesian machine learning methods can be used to learn the parameters, and we omit details for brevity.

Maximization in the iterative algorithm used Simulated Annealing to compute the most likely values for both the creation times and lineages. The Metropolis--Hastings~\cite{metropolis1953equation} algorithm was used to infer the distribution of creation times for the binaries.

We tested the lineage method on both synthetic data and real malware generated by MIT Lincoln Lab (MITLL) as part of the DARPA Cyber Genome effort~\cite{darpa2010}. The synthetic data was primarily used to compute the accuracy of the creation time learning and estimation, and our results showed that our probabilistic model is an effective means to capture this information. More importantly, we also used the synthetic data to verify if the iterative algorithm can improve the creation time estimation (i.e., is there benefit to joint inference). In Fig.~\ref{LineageTimeImprove} we show the reduction in the creation time error (expected creation time compared to real creation time) after lineage is performed, where we varied the obfuscation of the malware time stamps from  0\% to 100\%. As can be seen, on all but one of tests, our estimate of the creation times improves after the iterative lineage algorithm has run, demonstrating the value of joint inference on this problem.

We tested the lineage estimation algorithm on 17 lineages generated by MITLL. Fig.~\ref{LineageResults} the accuracy of the lineage algorithm on four metrics. We show the results of lineage generation on the same 17 families using our original implementation that did not use PP or joint inference (it computed lineage using a single estimate of the creation time). As can be seen, our lineage estimation algorithm can reconstruct lineages of malware with fairly high accuracy, and is a significant improvement over the non--PP solution. These results demonstrate that our probabilistic formulation of the lineage problem is an effective solution to construct lineages in light of uncertain and obfuscated data.

\section*{Acknowledgments}

This work was supported by the Defense Advanced Research Projects Agency (DARPA) under US Air Force contracts FA8750-10-C-0171 and FA8750-14-C-0011, with thanks to Mr. Timothy Fraser. The views expressed are those of the authors and do not reflect the official policy or position of the Department of Defense or the U.S. Government. Approved for Public Release, Distribution Unlimited.

%\subsubsection*{References}

\bibliographystyle{abbrv}
\bibliography{MAAGI_aij}

\begin{thebibliography}{1}

\bibitem{darpa2010}
DARPA.
\newblock Cyber defense (cyber genome) program, 2010.

\bibitem{dempster1977maximum}
A.~P. Dempster, N.~M. Laird, and D.~B. Rubin.
\newblock Maximum likelihood from incomplete data via the em algorithm.
\newblock {\em Journal of the Royal Statistical Society. Series B
  (Methodological)}, pages 1--38, 1977.

\bibitem{lakhotia_fast_2013}
A.~Lakhotia, M.~D. Preda, and R.~Giacobazzi.
\newblock Fast location of similar code fragments using semantic 'juice'.
\newblock In {\em Proceedings of the 2nd {ACM} {SIGPLAN} Program Protection and
  Reverse Engineering Workshop}, {PPREW} '13, page 5:1–5:6, New York, {NY},
  {USA}, 2013. {ACM}.

\bibitem{metropolis1953equation}
N.~Metropolis, A.~W. Rosenbluth, M.~N. Rosenbluth, A.~H. Teller, and E.~Teller.
\newblock Equation of state calculations by fast computing machines.
\newblock {\em The journal of chemical physics}, 21(6):1087--1092, 1953.

\bibitem{figaro2012}
A.~Pfeffer.
\newblock Creating and manipulating probabilistic programs with {F}igaro.
\newblock In {\em 2nd International Workshop on Statistical Relational AI},
  2012.

\end{thebibliography}

\end{document}